\documentstyle[graphicx,12pt]{article}
\renewenvironment{abstract}{\small \noindent{\bf Abstract:}\ }{}
\newcommand{\simgeq}{\; \raisebox{-0.4ex}{\tiny$\stackrel
{{\textstyle>}}{\sim}$}\;}
\newcommand{\simleq}{\; \raisebox{-0.4ex}{\tiny$\stackrel
{{\textstyle<}}{\sim}$}\;}
\newcommand{\beq}{\begin{equation}}
\newcommand{\beqar}{\begin{eqnarray}}
\newcommand{\eeq}[1]{\label{#1} \end{equation}}
\newcommand{\eeqar}[1]{\label{#1} \end{eqnarray}}
\newcommand{\eps}{\varepsilon}
\newcommand{\up}[2]{$^{#1#2}$}
\newcommand{\Qspec}{Q_{\rm spec}}
\newcommand{\figheight}{7cm}
\input{psfig}

\begin{document}

%\begin{flushright}
%Lund-MPh-00/03
%\end{flushright}

\title{\vspace{-1cm}
Rotational structure of $T$=0 and $T$=1 bands\\ 
in the $N$=$Z$ nucleus $^{62}$Ga}

\author{\bf Andrius Juodagalvis and Sven \AA berg\\
  \footnotesize
  Mathematical Physics Division, Lund Institute of Technology, \vspace{-0.2cm}\\
 \footnotesize P.O.\ Box 118, S-221 00 Lund, Sweden %
}
\date{}

\maketitle

\vspace{-7cm}
\begin{flushright}
Lund-MPh-00/03
\end{flushright}
\vspace{5cm}

\begin{center}
\begin{minipage}[t]{0.88\textwidth}

\begin{abstract}
The rotational behavior of $T$=0 and $T$=1 bands in the odd-odd
$N$=$Z$ nucleus $^{62}$Ga is studied theoretically using the spherical
shell model (laboratory frame) and the cranked Nilsson-Strutinsky
model (intrinsic frame). Both models give a good description of
available experimental data. The role of isoscalar and isovector
pairing in the $T$=0 and $T$=1 bands as functions of angular momentum
is studied in the shell model. The observed backbending in the $T$=0
band is interpreted as an unpaired band-crossing between two
configurations with different deformation. The two configurations
differ by 2p-2h and are found to terminate the rotational properties
at $I^{\pi}$=9$^+$ and $I^{\pi}$=17$^+$, respectively. $E2$-decay
matrix elements and spectroscopic quadrupole moments are
calculated. From the CNS calculation, supported by shell model
results, it is suggested that the low spin parts of the bands with
$T$=0 and $T$=1 correspond to triaxially deformed states with the
rotation taking place around the shortest axis (positive $\gamma$) and
intermediate axis (negative $\gamma$), respectively.  At lower spins
the configuration space $pf_{5/2}g_{9/2}$, used in the shell model
calculation, is found sufficient while also $f_{7/2}$ becomes
important above the backbending.
\end{abstract}

\vspace{0.3cm}

\footnotesize
\noindent
{\em Key words:\/} Shell model; Cranked Nilsson-Strutinsky; Rotational
bands; Pairing energy; %
%                \linebreak 
Quadrupole properties; N=Z; $^{62}$Ga.\par
\noindent
{\em PACS:\/} 21.10.G, 21.10.-k, 21.10.Ky, 21.60.C, 21.60.Cs
\end{minipage}
\end{center}

\section{Introduction}
One of the currently most interesting questions in nuclear structure
research concerns the interplay between isovector ($T$=1) and
isoscalar ($T$=0) neutron-proton pairing, see e.g.\ \cite{Mar99} and
references therein.  Effects from this interplay can be seen mainly in
nuclei with $N \approx Z$, since the $pp$ and $nn$ $T$=1 pairing
forces dominate in nuclei with proton (or neutron) excess. Odd-odd
nuclei with an equal number of neutrons and protons are of particular
interest since the isoscalar and isovector $np$ forces compete, and
the position of $T$=0 states relative to $T$=1 states is found to
change with the mass number: Most odd-odd $N$=$Z$ nuclei with $A \leq
40$ have $T$=0 ($I > 0$) ground states while most of the heavier nuclei
have $T$=1 ($I$=0) ground states.

Recently, several new states of positive parity were
found in the odd-odd $N$=$Z$ nucleus $^{62}_{31}$Ga$_{31}^{}$ 
\cite{Vincent98,St97}.
The states form a $T$=0 rotational band from $I$=1 up to $I$=17 
that backbends at $I \approx 9$, see fig.1. The $I$=1 $T$=0 state
decays to the $I$=0 $T$=1 ground state, and no excited $T$=1 states
were identified. The odd-odd nature of $^{62}$Ga implies blocking of
proton-proton and neutron-neutron isovector ($T$=1) pairing properties,
and both $T$=1 and $T$=0 neutron-proton pairing may be important \cite{Dean}.
From shell model calculations \cite{Vincent98} it seems that
the relative importance of these two pair fields
change as a function of angular momentum, since $T$=0
states are more favored than $T$=1 states for $I \simgeq 2$.
We shall address these questions by comparing experimental data to a shell
model (SM) calculation and an unpaired cranked Nilsson-Strutinsky 
(CNS) calculation.
\begin{figure}[tb]
\centerline{\psfig{figure=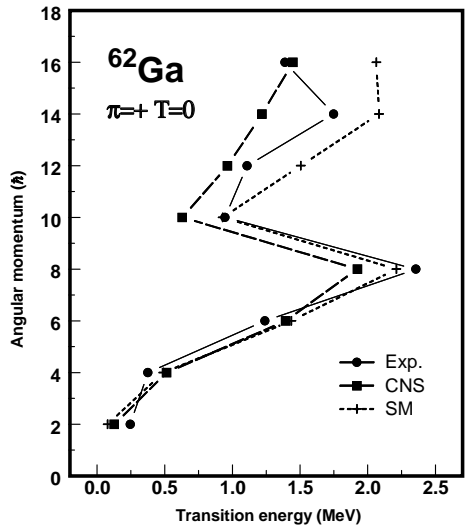,height=\figheight}}
\caption{Angular momentum versus transition energy for the $T$=0
positive-parity band in $^{62}$Ga. Filled circles connected
by straight lines show experimental values \cite{Vincent98,St97},
plus signs connected with short-dashed lines mark results from
a shell model calculation (discussed in section 2), 
and filled squares connected by
long-dashed lines show cranked Nilsson-Strutinsky results (section 3). 
Corresponding energies are shown in fig.2.}
\label{fig1}
\end{figure}

In the ground state of $^{62}_{31}$Ga$_{31}^{}$ the $f_{7/2}$ shell is
filled, and the three protons and three neutrons outside the $^{56}$Ni
core occupy the lowest-lying orbitals emerging from $p_{3/2}$ and
$f_{5/2}$. In the deformed shell model this implies a small prolate or
triaxial deformation ($\varepsilon \approx 0.17$) and thus collective
rotation.  The relatively small number of active particles allows
extended shell model calculations utilizing the valence shells
$pf_{5/2}g_{9/2}$. Collective rotation and backbending can then be
studied and compared in the laboratory frame (spherical shell model) and
in the intrinsic frame (deformed mean field) description. Similar
studies have recently been performed for the even-even nuclei
$^{48}$Cr \cite{Ca95,Andrius99} and \up36Ar \cite{Ar36}.

Results from a shell model calculation for $^{62}$Ga in the valence
space
$pf_{5/2}g_{9/2}$, utilizing a slightly modified version of the
effective interaction \cite{Nowacki}, are presented in section
2. Low-lying $T$=0 and $T$=1 bands are calculated, and we discuss the
role of isovector and isoscalar pairing for the different bands as a
function of angular momentum.  In section 3, low-lying $\alpha$=1 (odd
spins) and $\alpha$=0 (even spins) bands in $^{62}$Ga calculated in
CNS neglecting pairing are presented, and it is discussed how the
bands can be approximately identified with $T$=0 and $T$=1 states. In
section 4 the CNS and SM results are compared to
experimental energies.  Also shell occupation numbers are calculated
and compared in both models, and the nature of the observed
backbending is subsequently discussed.  In section 5 we report on
calculations of $B(E2)$-values and spectroscopic quadrupole moments in
SM as well as in CNS.  Finally, a short summary of the results is
given in section 6.

% --------------------------------------

\section{Shell model calculation}
At present large scale shell model codes \cite{Caurier} 
can easily handle 6 particles in the model space $pf_{5/2}g_{9/2}$. For
$^{62}$Ga this implies an assumption of a closed $f_{7/2}$ shell that
is a good approximation for most of the states discussed here. However, 
later we will see that this assumption is rather severe for the description
of the higher spin states. 

The shell model calculation was performed using the effective
interaction derived from a realistic G matrix 
whose monopole part has been phenomenologically adjusted
\cite{Nowacki}. The interaction was previously used to describe e.g.\ 
$^{76}$Ge and $^{82}$Se \cite{GeSePaper}. The single-particle energies
\cite{Nowacki} were taken from the experimental spectrum of $^{57}$Ni:
$\varepsilon_{p_{3/2}}$=0.0, $\varepsilon_{f_{5/2}}$=0.77,
$\varepsilon_{p_{1/2}}$=1.113, $\varepsilon_{g_{9/2}}$=3.0 MeV
(estimated). Recently, the single-particle energy of the $g_{9/2}$
shell in $^{57}$Ni was experimentally established \cite{G9spe}. The
measured value, $\varepsilon_{g_{9/2}}$=3.7 MeV, is larger than
previously used value, 3.0 MeV. However, it was found that the
agreement of the calculated spectrum with the experimental energies in
$^{62}$Ga is better if the single-particle energy of the
$g_{9/2}$-shell is lowered, thus we used $\varepsilon_{g_{9/2}}$=2.5
MeV, and this was the only modification of the interaction
\cite{Nowacki}.

% --------------------------------------

\subsection{$T$=0 and $T$=1 bands}
The $T$=0 band was observed up to $I$=17 \cite{Vincent98,St97} while
only one $T$=1 state, the $I$=0 ground state, was observed. The $T$=0
states form a rotational band that  backbends at $I$=9, see
fig.1.  The 246 keV, 376 keV and 1241 keV $\gamma$-rays were
assigned experimentally as corresponding to 
stretched $E2$ transitions, and the $T$=0
levels $1^+$, $3^+$, $5^+$ and $7^+$ could thus be assigned
\cite{Vincent98}, while the observed levels above the backbending
could be assigned \cite{St97,Dirk-private} mainly from energy
systematics and comparisons to SM calculations. No $B(E2)$ values have
so far been measured, except for the $3^+$ state that was found to be
isomeric with a life-time of 4.6(1.6) ns corresponding to $B(E2)=197(69)$
e$^2$fm$^4$ \cite{Vincent98}.

Experimental data is compared to calculated energies of low-lying
$T$=0 and $T$=1 states in fig.2.  Measured
$T$=1 energies above the $I$=0 state are taken from the analogue
states in $^{62}$Zn \cite{Zn62}. In agreement with data the
ground state is found to have $I$=0 and isospin $T$=1 in the SM
calculation. Also the excitation energy from the $T$=1 ground state to
the first excited $T$=0 $I$=1 state is well reproduced. With
increasing angular momentum the $T$=0 states become more favored, and
the $T$=0 band crosses the $T$=1 band already at $I \approx 2$. At low
spins the agreement with data is very good for the $T$=0 states while
the higher spin states ($I>11$) are less well reproduced. This is
presumably due to the restricted model space used in the SM
calculation, as will be discussed in sect.4. Fig.2a shows two 
calculated states
at spin $I$=9. The higher-lying state is connected to the yrast states
with lower spin by a strong $B(E2)$ value, and is constructed only
from $fp$ shell orbits. The configuration of the yrast $I$=9 state
involves the excitation of about two particles into the $g_{9/2}$
shell and is connected by a strong $E2$ matrix element to the yrast
$I$=11 state.  There is a strong $B(E2)$ transition from the yrast
9$^+$ to the yrare 7$^+$ state that also has two particles excited to
the $g_{9/2}$ shell. This state, however, is calculated to have larger
excitation energy than $9_1^+$ state. Thus in the sequence of odd spin
states with $T$=0 we see a change of configuration at $I\approx 9$
that later will be discussed in a greater detail. A similar configuration
change appears at $I\approx10$ in the $T$=1 band, see fig.2c. In the
$T$=1 band constructed from $fp$-shell orbits only, the energies of states
above spin $I$=4 are not very well reproduced in the SM calculation,
while the calculated energies of the $I$=10, 12, 14 and 16 states,
which involve the excitation of two particles to the $g_{9/2}$ shell, come
out quite close to the experimental values.
\begin{figure}[t]
\centering
\begin{minipage}[t]{0.5\linewidth}
\centerline{\psfig{figure=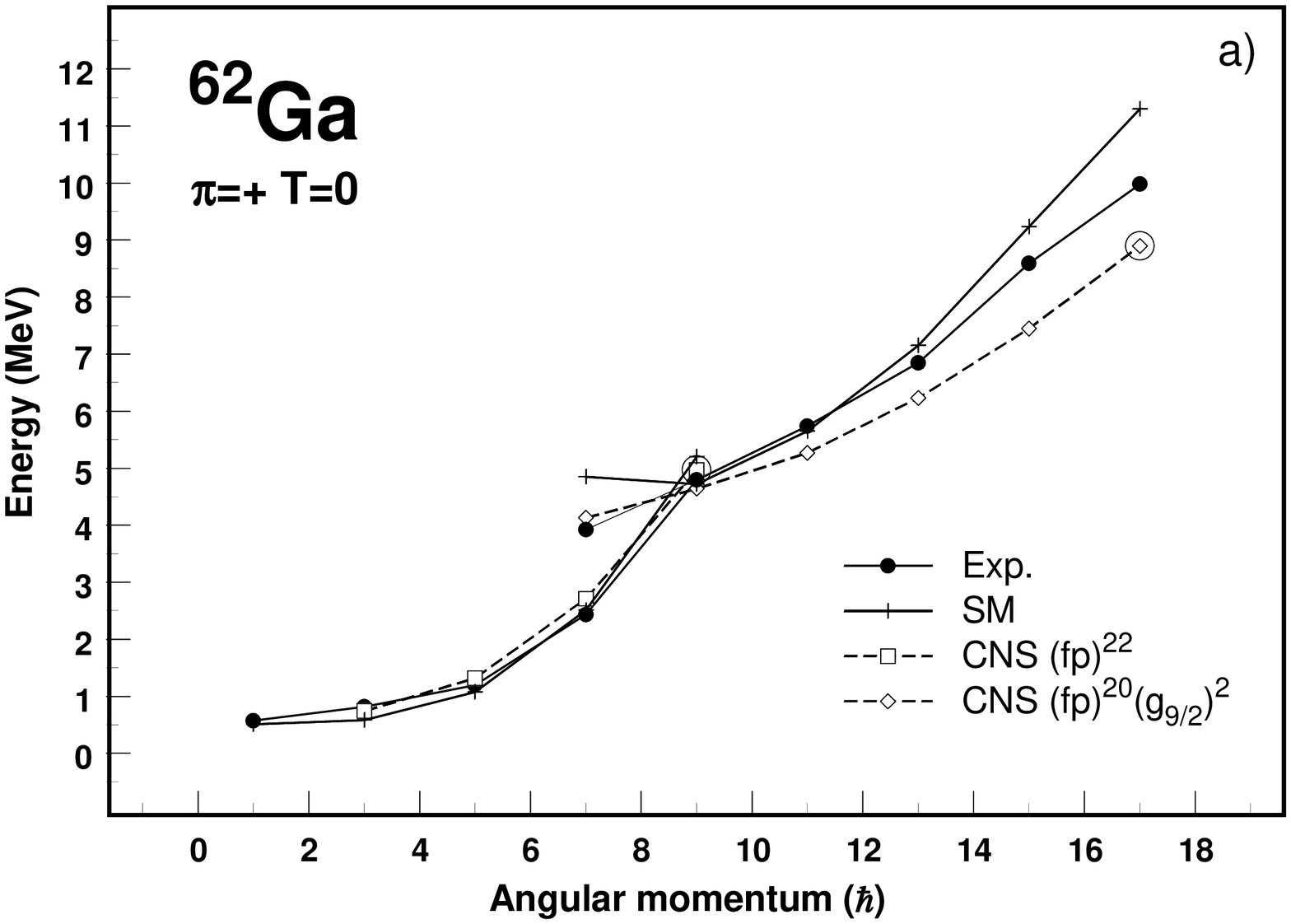,width=0.98\textwidth}}
\end{minipage}%
\begin{minipage}[t]{0.5\linewidth}
\centerline{\psfig{figure=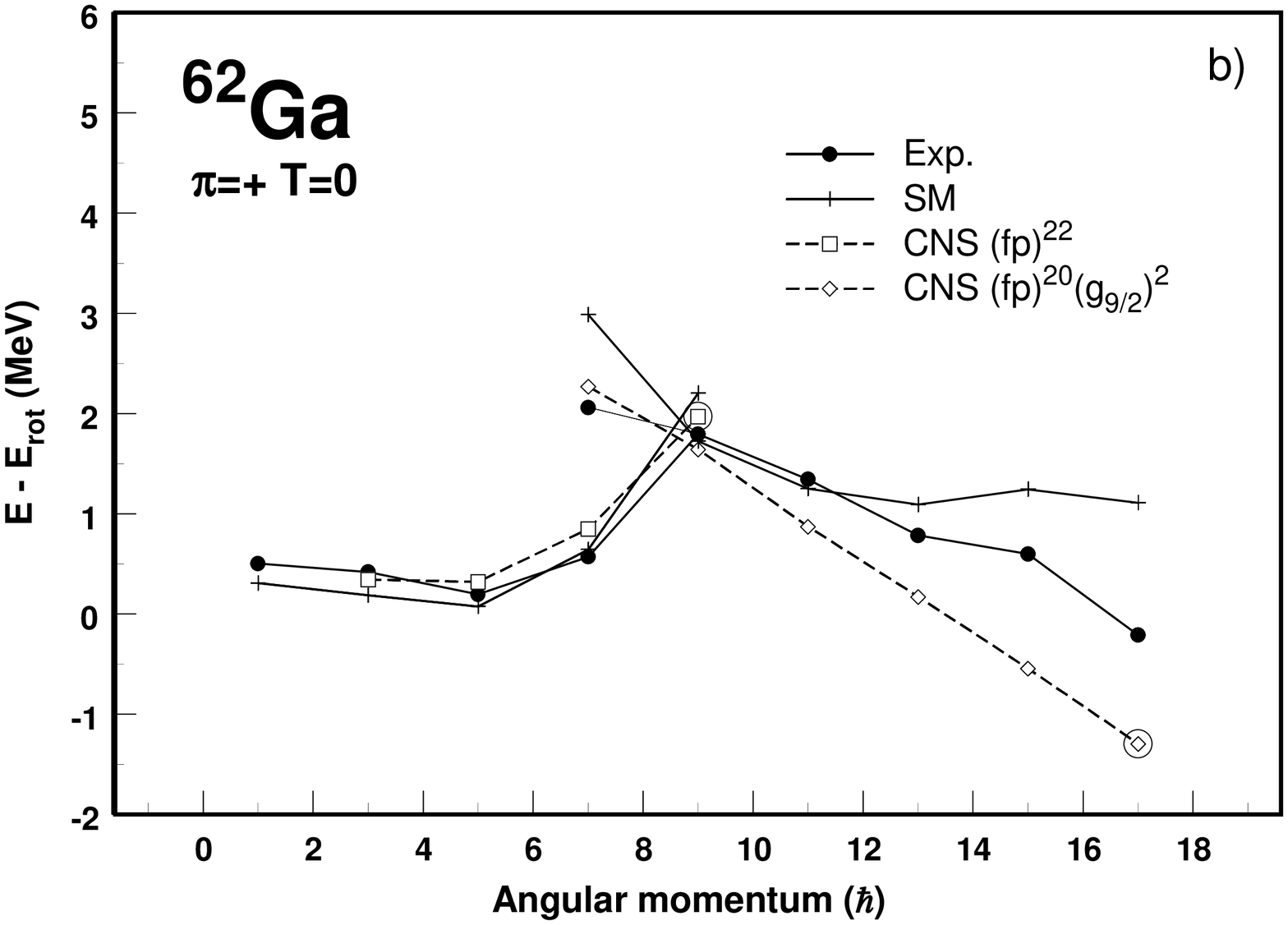,width=0.98\textwidth}}
\end{minipage}
\begin{minipage}[t]{0.5\linewidth}
\centerline{\psfig{figure=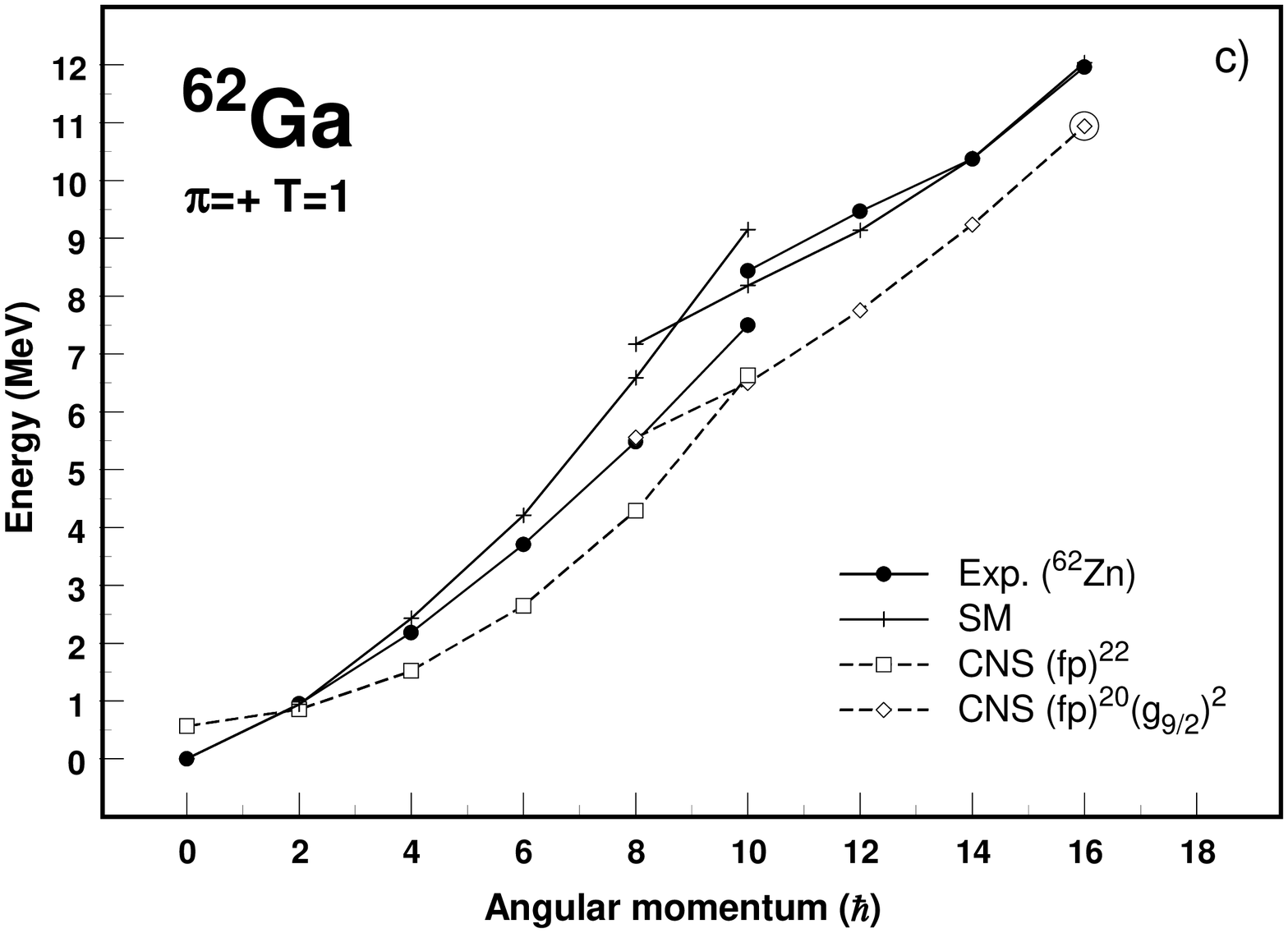,width=0.98\textwidth}}
\end{minipage}%
\begin{minipage}[t]{0.5\linewidth}
\centerline{\psfig{figure=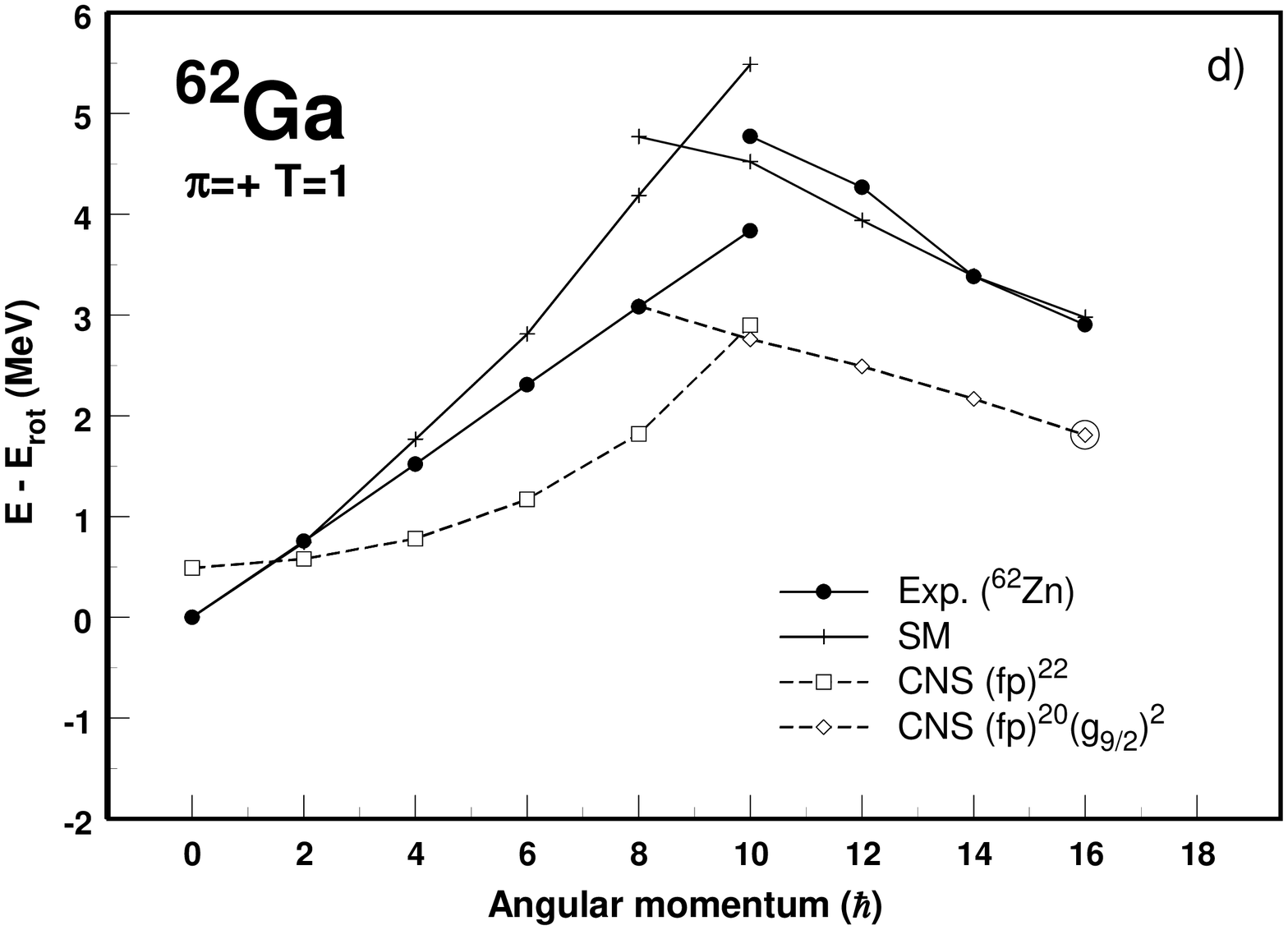,width=0.98\textwidth}}
\end{minipage}
\caption{Calculated and measured energies for (a) $T$=0 and (c) $T$=1
yrast states in $^{62}$Ga. Measured energies \cite{Vincent98,St97} are
shown by filled circles connected by solid lines, and shell model (SM)
results (model space $pf_{5/2}g_{9/2}$) by plus signs.  Measured $T$=1
energies above the $I$=0 state are taken from the analogue states in
$^{62}$Zn \cite{Zn62}.  The observed state at 3924 keV
\cite{St97} has been assigned as a $T$=0 7$_2^+$ state
\cite{Dirk-private}, and connected to the 9$_1^+$ state by a thin
solid line. The dashed lines show energies calculated in the cranked
Nilsson-Strutinsky (CNS) model without pairing; open squares and open
diamonds imply configuration $(fp)^{22}$ and $(fp)^{20}g_{9/2}^2$,
respectively. In CNS the $T$=0 bands terminate at $I$=9 and $I$=17,
respectively, that is marked by open circles.  CNS calculated yrast
states with $\alpha$=0 (cf.\ \cite{Zn62}) are compared to $T$=1
states in (c). In (b) and (d) a rotational reference of $E_{\rm
rot}=0.033I(I+1)$ MeV is subtracted.}
\label{fig2}
\end{figure}

The increased favoring of $T$=0 states relatively to
$T$=1 states implies a moment of inertia that is 
larger for the $T$=0 band than for the $T$=1 band.
The origin of this difference may be found in different pairing properties
of the two bands. We shall therefore study $T$=0 and $T$=1 pairing energy 
contributions to different states in the $T$=1 and $T$=0 bands.

% --------------------------------------

\subsection{Pairing in the SM}
The pairing energy in the shell model is calculated by taking energy
difference of states calculated using the full Hamiltonian and the
Hamiltonian from which a schematic $L$=0 pairing interaction is
subtracted (cf.\ \cite{Cr-pairing}). The normalized
form of the pairing interaction was used \cite{DufourZuker} with the
strength parameters ${\bar G}_{01}$=2.98 and ${\bar G}_{10}$=4.75 for
the $T$=1 and $T$=0 pairing, respectively. Since the excited $T$=0
band has a very different configuration as compared to the lower $T$=0
band, it was possible to calculate the pairing energy for each
configuration separately even in the band-crossing region.

In fig.3 $T$=0 and $T$=1 pairing energy contributions in the orbital
angular momentum $L$=0 channel is shown for $T$=0 odd spin states (fig.3a),
for $T$=1 even spin states (fig.3b), and for $T$=0 even spin states (fig.3c).
The $T$=1 $I$=0 ground state is seen to have 
approximately equal amounts of pairing
energy contributions from the $T$=0 and $T$=1 channels. For the lowest 
$T$=0 state ($I$=1), on the other hand, the 
isoscalar neutron-proton ($T$=0) pairing plays a larger role than the
isovector ($T$=1) pairing, although the latter 
has contributions from neutron-neutron, 
proton-proton and neutron-proton pairing. The same is valid for the
lowest $T$=0 state with even spin, $I$=2.
\begin{figure}[tb]
\centerline{\psfig{figure=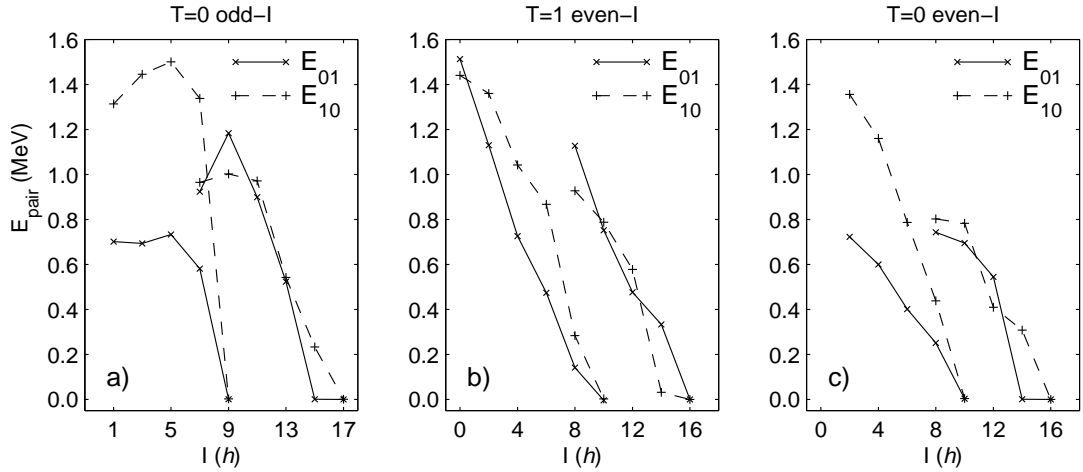,width=0.9\textwidth}}
\caption{Pairing energies from the SM calculation for (a) $T$=0 
odd spin states, (b) $T$=1 even spin states, and (c) $T$=0 even spin
states. Solid lines are used for $T$=1 pairing ($nn$+$pp$+$np$) in the
$J$=$S$=0 channel, E$_{01}$, and dashed lines for $T$=0 pairing ($np$)
in the $J$=$S$=1 channel, E$_{10}$.  The two configurations,
$(fp)^{22}$ and $(fp)^{20}g_{9/2}^2$, are yrast between
0$\leq$$I$$\leq$8 and 9$\leq$$I$$\leq$17, respectively.}
\label{fig3}
\end{figure}

Due to isospin symmetry all parts of the isovector pairing, $nn$, $pp$
and $np$, are identical in the $T$=0 states. The odd-odd nature of
$^{62}$Ga implies that at least one isovector pair is broken in the
$T$=0 states, i.e.\ the isovector pairing is reduced. From fig.3a it
is thus seen that each component of the isovector pairing mode
contributes to the $T$=0 ($I$=1) state only about 0.25 MeV.

The isoscalar pairing energy is calculated as about 1.4 MeV in lowest $T$=0 
as well as in lowest $T$=1 states, while the total isovector pairing 
energy contributions are about 0.7 MeV and 1.5 MeV, respectively.
In total, the $T$=1 ($I$=0) ground state thus gets a pairing energy 
contribution of about 2.9 MeV (fig.3b), while the corresponding energy is 
about 2.1 MeV for the lowest $T$=0 state ($I$=1), see fig.3a. 
This favoring by about 0.8 MeV pairing energy of the lowest $T$=1 state 
compared to lowest $T$=0 state plays an important role in 
making the $T$=1 state to become the ground state. 

With increasing spin the pairing energy, both isoscalar and isovector,
is approximately constant for the lowest odd-$I$, $T$=0 states, while
for the $T$=1 states pairing energy decreases rather drastically,
especially the isovector part. That is, the moment of inertia for the
odd-$I$ $T$=0 states is more or less unaffected by pairing
correlations (for $I<9$), while it is reduced by about 50\% for the
$T$=1 states. This causes the $T$=1 band to cross the $T$=0 band
already at $I \approx 2$.  The pairing energy decreases with
increasing spin also for even-$I$ $T$=0 states, see fig.3c. However,
since $T$=1 pairing is partly blocked the effect is less drastical for
these states as compared to $T$=1 states.

In SM calculations of the ground band in 
$^{48}_{24}$Cr$^{}_{24}$ it was found that isoscalar pairing correlations 
are less dependent on increasing spin than isovector pairing correlations 
\cite{Cr-pairing}, i.e.\ the moment of inertia is less sensitive to
isoscalar pairing. The present results for $^{62}$Ga do, however,  
suggest similar energy correlations with increasing spin from isoscalar 
and isovector pairing, or even a larger energy change due to isoscalar 
pairing, see fig.3. The different pairing behavior found in the two nuclei 
is explained by the size of the pairing energy in the ground state.
In $^{48}$Cr isovector pairing gives a much larger energy 
contribution to the $I$=0 ground state, $E_{01}$= 3.6 MeV,  
as compared to isoscalar pairing, $E_{10}$=2.0 MeV \cite{Cr-pairing}. 
In $^{62}$Ga, on the other hand, the two
contributions are similar for the $T$=1 ground state ($E_{01}\approx E_{10}
\approx 1.5$ MeV), while isoscalar
pairing energy is largest for $T$=0 states. Since isoscalar and isovector
pairing energy both decrease with increasing spin, and are 
approximately zero at the same spin value
(when seniority is large enough), 
the energy change with spin is mainly determined
by the energy contribution to the starting spin state.
However, the pairing energy may decrease in different ways for
different bands, see fig.3.

The configuration of the low-spin states can be followed smoothly to the
non-yrast $I$=9 and $I$=10 states (for $T$=0 and $T$=1).
These states constitute
approximately maximum seniority states ($v=6$),
and all studied pairing energy contributions vanish there.
The yrast states with spin values higher or equal to $I$=9 (or $I$=10) 
correspond to states where approximately two particles are excited to
the $g_{9/2}$ shell, i.e.\ they have the configuration
$(pf_{5/2})^{4}g_{9/2}^2$ (or, if also the particles in the filled
$f_{7/2}$ shell are counted, $(fp)^{20}g_{9/2}^2$).
For these states pairing correlations are again active, see fig.3. 
In a similar way as for the $(fp)^{22}$ configuration,
isoscalar and isovector pairing decrease with increasing spin 
and are zero for the $I$=17 and $I$=16 states, which are the maximum  
spin states in the $(pf_{5/2})^{4}g_{9/2}^2$ configuration.
The main difference between the 
two configurations, $(pf_{5/2})^{6}$ and $(pf_{5/2})^{4}g_{9/2}^2$, 
is that the isovector pairing
is of a similar size as the isoscalar pairing in the $(fp)^{20}g_{9/2}^2$
configuration. This may be explained in the following way.
Since the two odd nucleons excited to the $g_{9/2}$ shell
can couple to isospin zero (preferentially with $I$=9), 
the two pairs in the $fp$ shell 
are fully active for $T$=0 pairing as well as for $T$=1 pairing.

% --------------------------------------

\subsection{High-spin behavior}
The somewhat irregular energy behavior of the $(fp)^{20}g_{9/2}^2$ band 
between $I$=9 and 17, as seen in experimental data (fig.2), 
can be understood from a sche\-ma\-tic seniority model
discussed in \cite{Ju98}. Two neutrons and two protons
occupy $(p_{3/2},f_{5/2},p_{1/2})$ and one neutron and one proton 
occupy $g_{9/2}$.
The lowest energy of a given spin is obtained by first minimizing
the total seniority, $v=v_p+v_n$, and then maximizing the reduced isospin,
$t=\frac{1}{2}|v_p-v_n|$, where $v_p$ and $v_n$ are the proton and neutron
seniority, respectively. With one neutron and one proton into the $g_{9/2}$
shell,
$9^+$ is the highest spin that can be constructed for 
$v$=2. By breaking the neutron (or proton) pair in 
$(p_{3/2},f_{5/2},p_{1/2})$ at most 4 more units of $\hbar$ can be 
obtained, and
the $11^+$ and $13^+$ states correspond to $v$=4 and $t$=1 in this 
simple model. In the same
way the remaining proton (or neutron) pair is broken giving $15^+$ and $17^+$
with $v$=6 and $t$=0. The arc-like structure seen in the energy sequence 
$I=9^+ - 13^+$, and much more evident for $I=13^+ - 17^+$, thus reminds of
a seniority coupling scheme in $(p_{3/2},f_{5/2},p_{1/2})$. 
The smaller collectivity (smaller configuration mixing) 
for the higher spin states implies a coupling scheme
more similar to a pure seniority scheme. The band termination at $17^+$ is
thus smooth ($\Delta v$=0 and $\Delta t$=0) and favored \cite{Ju98}, and
is very similar to the band terminations seen e.g.\ at $8^+$ in 
$^{60}_{30}$Zn$_{30}^{}$ ($(p_{3/2}f_{5/2})^4$), 
at $12^+$ in $^{44}_{22}$Ti$_{22}^{}$ 
($f_{7/2}^4$) and at $15^+$ in $^{46}_{23}$V$_{23}^{}$ ($f_{7/2}^6$). 
However, as will be shown in next section,
the CNS calculation suggests that also the $f_{7/2}$
shell is important for the description of the higher spin states in $^{62}$Ga, 
and the band termination scenario may be more complicated than 
discussed here.

% --------------------------------------

\section{Cranked Nilsson-Strutinsky calculation}
The cranked Nilsson-Strutinsky calculation is performed with standard
parameters \cite{Ragnarsson}, and with the possibility to fix
the configuration in the minimization procedure \cite{CNSFixConf}.  We
allow for a free minimization of the two quadrupole degrees of
freedom, $\eps$ and $\gamma$, as well as one hexadecapole degree of
freedom, $\eps_4$.  All kinds of pairing interactions are
neglected. It is the experience that also unpaired CNS calculations
give a good insight into the nuclear structure, in particular at
higher spins.  Below we present the results of a CNS calculation of
the rotational behavior of $^{62}$Ga. Since SM states are additionally
classified with the isospin quantum number, we first need an
understanding of isospin in the CNS model (subsection 3.1). In
subsection 3.2 we discuss the behavior of the lowest rotational bands with
even as well as odd spins, i.e., with signature quantum numbers
$\alpha$=0 and $\alpha$=1, respectively.

% --------------------------------------

\subsection{Isospin in CNS}
In the cranked Nilsson-Strutinsky model isospin is not a good quantum
number.  The Coulomb force (included in the present model through
different sets of Nilsson parameters for protons and neutrons) breaks
isospin, and total wave-functions with good isospin are not
constructed.  If pairing is neglected one may still construct states
of approximately good isospin for nuclei with $N \approx Z$ (cf.\
\cite{TinCNS}). For odd-odd $N$=$Z$ nuclei we may approximately
identify a configuration with isospin $T$=0 if it cannot be realized
(due to the Pauli principle) in the even-even neighbor. That is, if
all protons are placed in ``exactly'' the same single-particle orbits
as the neutrons (the orbits are nearly the same since the Coulomb
force plays a minor role), the configuration is identified with
isospin $T$=0.  This means that the two odd nucleons in $^{62}$Ga have
the same signature and parity, i.e.\ they form a pair with the total
signature $\alpha$=1 and parity $\pi$=+. The same applies for each
neutron-proton pair, and due to an odd number of pairs the total
signature and parity is thus $\alpha$=1 and $\pi$=+, corresponding to
a positive parity rotational band with odd spins.  The shell filling
of the three least bound protons and neutrons in $^{62}$Ga for the
$\alpha$=1 and $\alpha$=0 bands at appropriate equilibrium
deformation, is shown in fig.4. On the right-hand side (positive
$\gamma$) we show filling of proton-neutron pairs in (approximately)
identical orbitals, leading to the total signature $\alpha$=1 which
may be identified with an isospin $T$=0 configuration.
\begin{figure}[tb]
\centerline{\psfig{figure=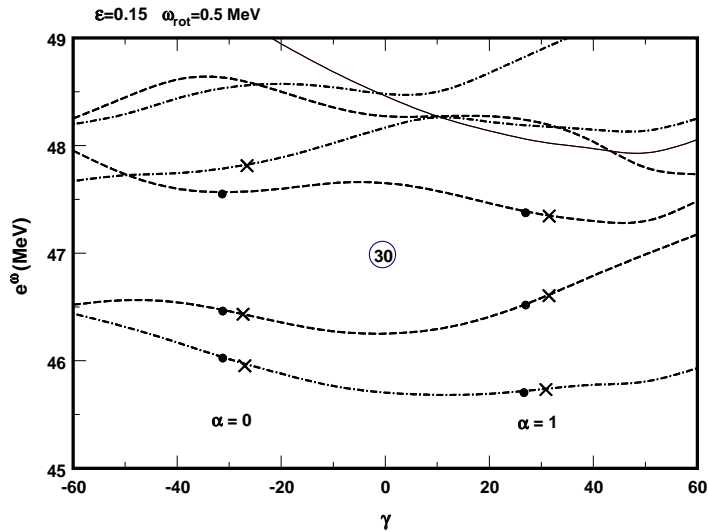,height=\figheight}}
\caption{Single-particle energies in CNS versus $\gamma$ for $\varepsilon$=
0.15 and $\omega_{\rm rot}$=0.5 MeV ($I$$\approx$6$\hbar$),
valid for neutrons as well as for protons (approximately) in $^{62}$Ga. 
The deformation 
$\gamma=-60^{\circ}$ corresponds to an oblate shape with collective rotation,
$\gamma=60^{\circ}$ corresponds to the same shape but with non-collective
rotation, and $\gamma=0^{\circ}$ to prolate shape with collective rotation.
The orbital with positive parity and signature $\alpha$=1/2 
(lowest $g_{9/2}$ orbital)
is drawn by a solid line, orbitals with negative parity are drawn with 
dashed lines ($\alpha$=1/2) and dot-dashed lines ($\alpha$=$-$1/2).
We show the filling of the $\alpha$=0 ($T$=0/$T$=1) configuration at 
negative $\gamma$
and the $\alpha$=1 ($T$=0) configuration at positive $\gamma$ (cf fig.5)
with the three least bound protons (solid circles) and neutrons (crosses).}
\label{fig4}
\end{figure}

If the two odd nucleons are placed in orbits with different
signatures, the proton (neutron) can be replaced by a neutron
(proton).  Consequently, the state can be realized also in the
even-even neighbors.  This situation is shown in the left-hand part of
the single-particle diagram (fig.4). The last filled neutron and
proton now occupy orbitals with different signatures, and a state with
total signature $\alpha$=0 is formed.  Since the orbits of the odd
neutron and proton can be interchanged, two almost degenerate bands
are formed in the CNS. Both bands have $\alpha$=0, i.e.\ even spins,
and form the basis of one $T$=0 band (signature partner of the
$\alpha$=1 $T$=0 band) and one $T$=1 band. Without additional
correlations between particles the two bands should thus come very
close in energy, and we may compare the calculated $\alpha$=0 band(s)
with the lowest-lying $T$=1 band as well as with the lowest $T$=0 band
with even spins.

With this simple classification, neglecting pair correlations,
the $T$=0, $\alpha$=1 band will in general be lowest in energy, and the 
excitation energy to the $T$=1 band approximately correspond to the 
calculated signature splitting of the last filled orbital. 
However, the two signature partners may correspond to nuclear shapes
with different deformations, implying a more complicated relation between the 
energies of the two bands also in unpaired calculations.

% --------------------------------------

\subsection{Total signature $\alpha$=0 and $\alpha$=1 bands}
The calculated yrast $\alpha$=0 (``$T$=1/$T$=0'') and $\alpha$=1 (``$T$=0'') 
band energies are shown in fig.2, and equilibrium
deformations are shown in fig.5. In figs.2b and 2d 
a rotational reference has been 
subtracted to facilitate the reading of the figure.
At low spin values the configuration with
all 22 particles outside $^{40}$Ca placed 
in the $fp$ shell, $(fp)^{22}$, comes out as lowest in energy
for even as well as for odd spin values in the CNS calculation.
At $I \simgeq 9$ it is more favorable to excite two particles to 
the $g_{9/2}$ shell and the configuration $(fp)^{20}g_{9/2}^2$ becomes 
yrast for states in the $\alpha$=0 as well as $\alpha$=1 bands.

All states in the yrast $(fp)^{22}$ configuration, the $\alpha$=1 band
as well the $\alpha$=0 band, are calculated to be triaxial with rather
large values of the triaxiality parameter $\gamma$, see fig.5. The
$\alpha$=1 states correspond to rotation around the smallest axis
(positive $\gamma$), while the $\alpha$=0 states correspond to
rotation around the (classically forbidden) intermediate axis
(negative $\gamma$).  The different types of triaxiality for the two
bands can be understood from the behavior of single-particle energies
as functions of triaxiality parameter, $\gamma$, see fig.4.  For the
$\alpha$=1 band the two odd nucleons occupy an orbit with the signature
$\alpha$=+1/2 which prefers positive $\gamma$, and the energy for the
total configuration obtains a minimum at $\gamma>0$.  In the
$\alpha$=0 states one of the two odd nucleons occupies an orbital with
$\alpha$=$-$1/2 which prefers negative $\gamma$, and the total
configuration ends up with a deformation $\gamma<0$.
\begin{figure}[tb]
\centerline{\psfig{figure=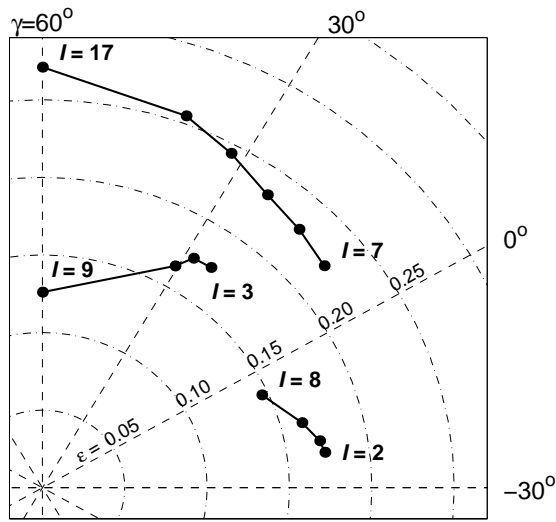,height=\figheight}}
\caption{Calculated equilibrium deformations in the 
($\eps$,$\gamma$)-plane for states in $^{62}$Ga. Bands with
spins $I^{\pi}$=3$^+$-9$^+$ 
and $I^{\pi}$=2$^+$-8$^+$ have the configuration  $(fp)^{22}$, and the
states $I^{\pi}$=7$^+$-17$^+$ have the configuration $(fp)^{20}g_{9/2}^2$. 
The two bands with odd spin values are seen to terminate at 
$\gamma$=60$^{\circ}$ (non-collective rotation) at 
$I$=9 and 17, respectively.}
\label{fig5}
\end{figure}
\begin{figure}[htb]
\begin{center}
\begin{minipage}[t]{0.49\textwidth}
\centerline{\psfig{figure=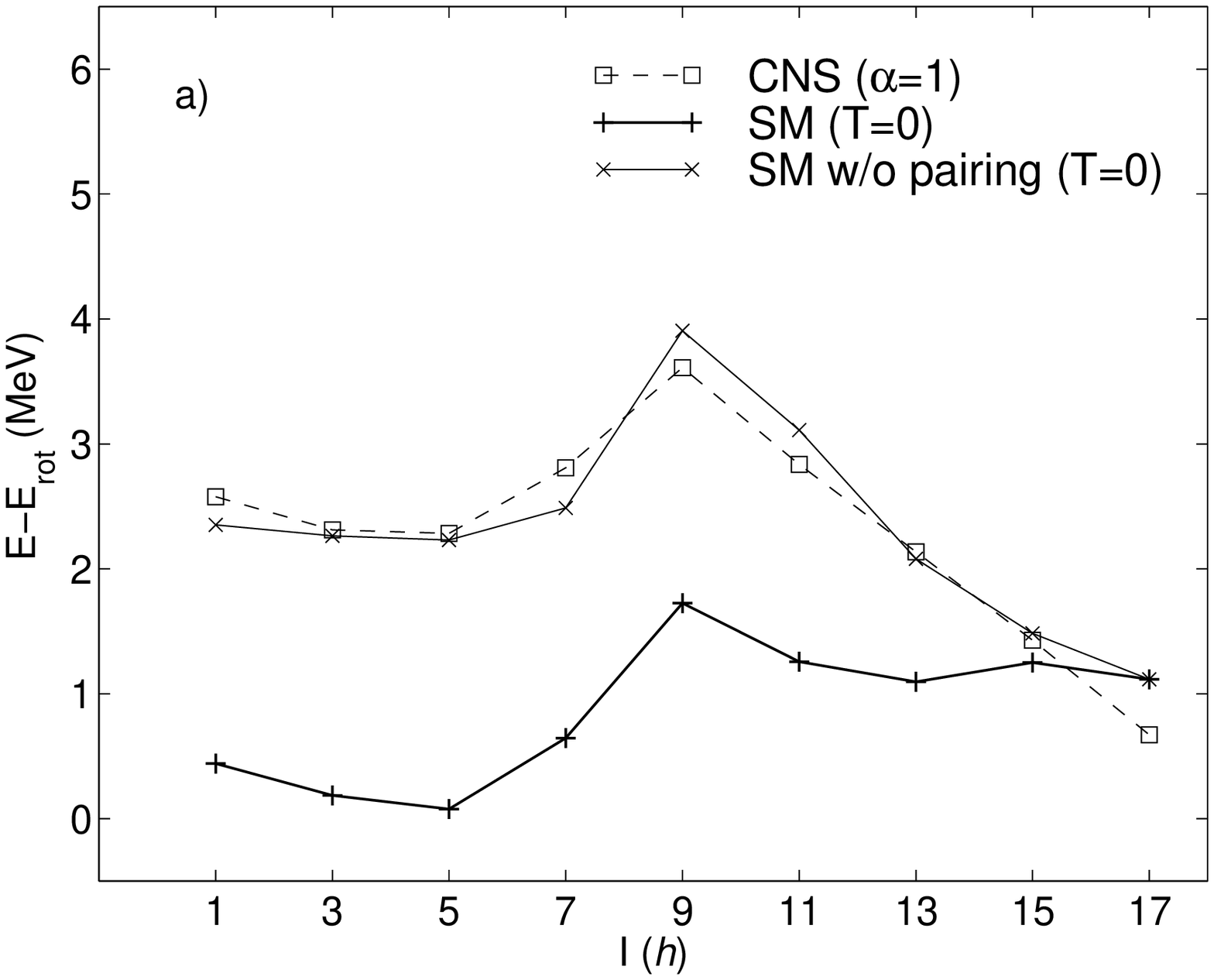,width=0.98\textwidth}}
\end{minipage}%
\begin{minipage}[t]{0.49\textwidth}
\centerline{\psfig{figure=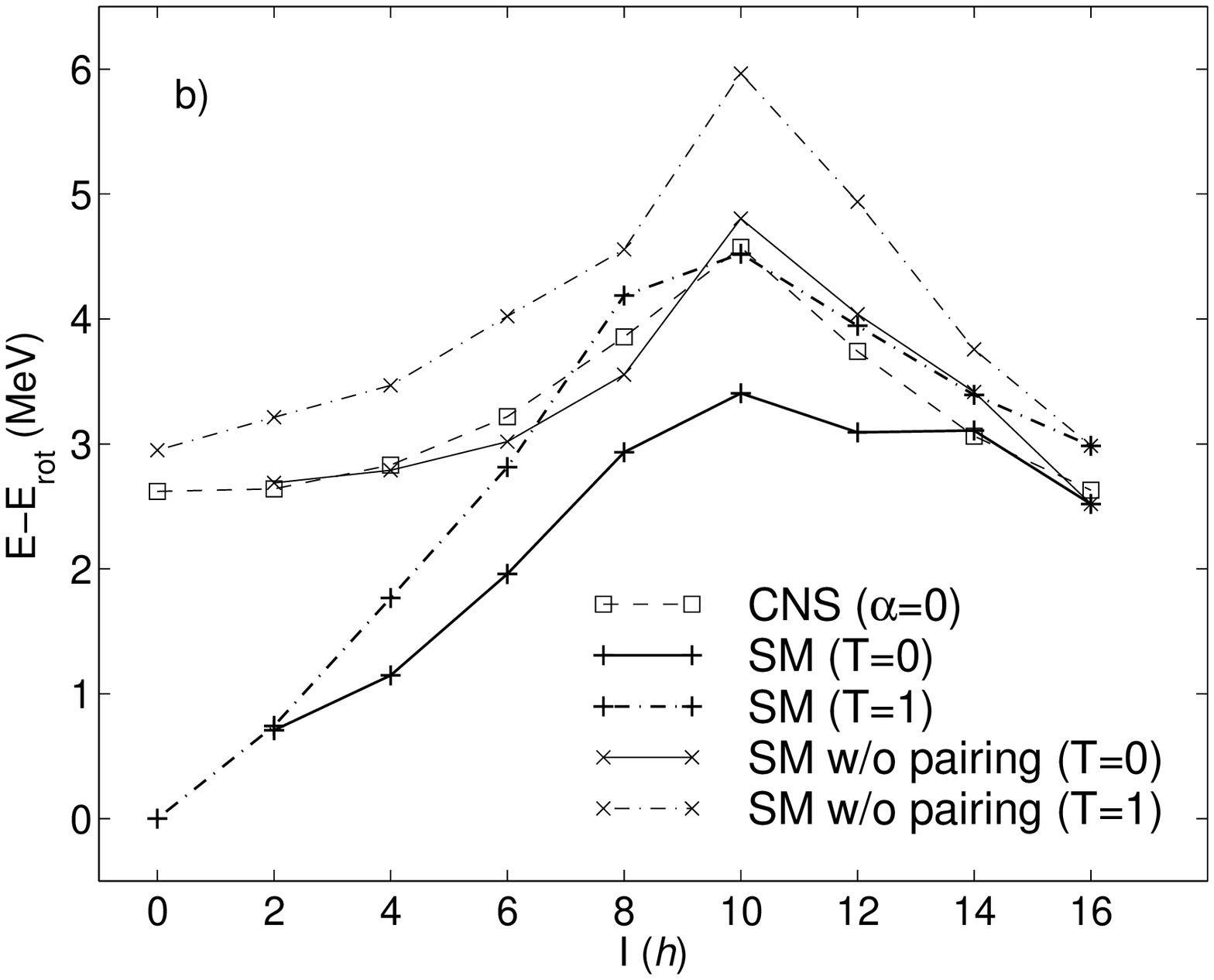,width=0.98\textwidth}}
\end{minipage}
\end{center}
\caption{Yrast energies in $^{62}$Ga minus a rotational reference (cf caption
to fig.2)
for (a) odd spin states in the $T$=0 band, and (b) even spin states 
in the $T$=1 and $T$=0 bands. Thick lines denote results from full SM 
calculation. The pairing energy coming from both $T$=0 ($S$=1) and 
$T$=1 ($S$=0) pairs (see text and fig.3) has been subtracted for the 
energies shown by thin lines. States of $T$=0 and $T$=1 character 
are connected by solid and dot-dashed lines, respectively.
For comparison we show unpaired CNS energies with dashed lines for yrast 
(a) $\alpha$=1 states and (b) $\alpha$=0 states. Compared to fig.2 the 
CNS energies have been arbitrarily shifted up by 2.5 MeV. 
}
\label{fig6}
\end{figure}
With increasing spin the $\gamma$ value 
increases for both the $\alpha$=1 and $\alpha$=0 bands; for 
the $\alpha$=1 band from $\gamma \approx 25^{\circ}$ for
the $I$=3 state to $\gamma = 60^{\circ}$ (i.e.\ band termination in a 
non-collective state, cf.\ \cite{BandTermReview}) at $I$=9, and
for the $\alpha$=0 band 
from $\gamma \approx -25^{\circ}$ for the $I$=2 state to 
$\gamma = -5^{\circ}$ for $I$=8,
see fig.5. The termination of the $\alpha$=1 band 
at $I^{\pi}$=9$^+$ corresponds to an oblate shape 
where the angular momentum components $\Omega$=3/2 and 1/2 of $p_{3/2}$
and the $\Omega$=5/2 component of $f_{5/2}$ are occupied by the 
three protons and three neutrons outside the $f_{7/2}$ shell. 
Electromagnetic properties of these two bands are discussed in section
5 below.

States with the configuration $(fp)^{20}g_{9/2}^2$, $\alpha$=1
(``$T$=0'') are also found to be triaxial with rotation around the
smallest axis (positive $\gamma$), see fig.5. The degree of
triaxiality increases smoothly along the band from $\gamma \approx
10^{\circ}$ at $I$=7 to the oblate symmetric state, $\gamma =
60^{\circ}$, at $I$=17 which terminates the rotational band within
this configuration.  The $\alpha$=0 (``$T$=0/$T$=1'')
$(fp)^{20}g_{9/2}^2$ band behaves in a similar way as the $\alpha$=1
$(fp)^{20}g_{9/2}^2$ configuration, and terminates at $I$=16. The CNS
calculated energies of $\alpha$=1 isobaric analog states in \up62Zn
were reported in \cite{Zn62}.

% --------------------------------------

\section{Comparison between unpaired CNS and SM}
In this section we compare results obtained in the two models. In
subsection 4.1 energies are compared, and occupation numbers
in spherical $j$-shells are compared in subsection 4.2. 
The backbending observed in the $T$=0 band 
(fig.1) is discussed in subsection 4.3.

% --------------------------------------

\subsection{Energies}
In fig.2 energies calculated in the CNS model are compared with both
SM results and experimental values: the $\alpha$=1 states in figs.2a
and 2b (experimental data from $^{62}$Ga), and the $\alpha$=0 states
in figs.2c and 2d (experimental data from $^{62}$Zn). In particular
the low-spin $\alpha$=1 states agree quite well with data for $T$=0
states (as well as with SM results), while the even spin states show
large deviations with the $T$=1 states. In fact, the moment of inertia
of the $T$=0 band appears very similar in unpaired CNS and SM
calculations (and in experimental data). This is since pairing
correlations are relatively stable for most $T$=0 states, while it is
strongly changing with increasing spin for $T$=1 states, as discussed
in sect.2.

Since the CNS calculation is performed without pairing it seems more
relevant to compare CNS energies with unpaired SM energies. This is
done by subtracting isovector and isoscalar ($L$=0) pairing
contributions, taken from fig.3, from the SM energies for each
individual state, and in fig.6 we compare energies of the lowest
states calculated in the CNS model with SM results. The energies are
indeed very similar in the unpaired CNS and the ``unpaired'' SM. Note,
however, that it is not straightforward that the removed pairing
energies in SM correspond to the (not included) pairing in CNS.
 
From fig.6a it is seen that the yrast sequence of odd spin states
($\alpha$=1), calculated in unpaired CNS, come quite close to the
lowest odd spin states in SM calculation ($T$=0 states) when pairing
has been removed. The fact that low-spin CNS energies agree with full SM
results (fig.2) as well as with SM results without pairing (fig.6a) is
explained by the rather constant pairing energy contribution for the
low spin, odd-$I$, $T$=0 states, see fig.3a.  However, the non-yrast
9$^+$ state, that terminates the $(fp)^{22}$ configuration has a
pairing energy contribution which is approximately zero. The SM
results without pairing thus deviate rather strongly from the CNS
results if all states in the $(fp)^{22}$ band are studied. The origin
of this deviation is not clear.

Since $\alpha$=0 states (even spins) calculated in CNS 
can be compared either to $T$=1 even spin states
or to $T$=0 even spin states, both types of SM states 
are shown in fig.6b. In the full SM calculation the rotational bands
with $T$=0 and $T$=1 even spin states show
quite different energy behavior. However, when pairing has been subtracted
they are similar, and also similar to unpaired CNS results.
The unpaired $T$=1 states come approximately 1 MeV higher in energy for most
spin states compared to the $T$=0 states, and the $T$=0 states are close
to unpaired CNS energies. 
The approximately constant energy difference between $T$=0 and 
$T$=1 states that remains in the SM calculation after all $L$=0 pairing
has been subtracted, originates from other parts of the interaction. 

% --------------------------------------

\subsection{Sub-shell occupancy}
A more detailed understanding of the structure of the $T$=0 and $T$=1
bands can be achieved by studying wave functions.  In SM the wave
functions are extremely complex and composed by very many
components. A simple view can, however, be obtained by studying
occupation numbers in different $j$ shells. Such studies are also
feasible in CNS, and some aspects of wave functions can then be
compared in the two models.

The added neutron and proton occupation numbers in spherical
$j$-shells for some selected states are compared in Table 1. The method
described in ref.\cite{Mg} was used to calculate the occupancies in
CNS. In SM, the total sum of occupation numbers in the model space
($p_{3/2}$, $f_{5/2}$, $p_{1/2}$ and $g_{9/2}$) is 6, while CNS does
not have such limitation for the configuration space. The $f_{7/2}$
shell belongs to the core in SM and is consequently always
``occupied'' by 16 particles.
\begin{table}[t]
\caption{
Total occupancies of nucleons (protons plus neutrons)
in the {\em spherical} $j$-shells
$f_{7/2}$, $p_{3/2}$, $f_{5/2}$, $p_{1/2}$ from the $N$=3 shell, and
$g_{9/2}$ from the $N$=4 shell, for a number of states calculated in CNS
and in SM. In the SM the valence space $pf_{5/2}g_{9/2}$ was used, and
the occupation number of $f_{7/2}$ is thus by default equal to 16.}
\vspace{1mm}

%\begin{center}
\resizebox{\textwidth}{!}{%
\begin{tabular}{||r|c|c||r|r|r|r|r||r|r|r|r|r||} \hline
\multicolumn{3}{||c||}{\em }  
& \multicolumn{5}{c||}{CNS} & \multicolumn{5}{c||}{SM} \\
\hline
$I^{\pi}$ & $T$ & conf. & $f_{7/2}$ & $p_{3/2}$ & $f_{5/2}$ & $p_{1/2}$ 
& $g_{9/2}$
& $f_{7/2}$ & $p_{3/2}$ & $f_{5/2}$ & $p_{1/2}$ 
& $g_{9/2}$ 
\\ \hline
3$^+$ & 0 & $(fp)^{22}$ & 15.50 & 3.83 & 1.77 & 0.67 & 0.08 
& 16 & 3.11 & 1.67 & 1.07 & 0.15 \\
9$^+$ & 0 & $(fp)^{22}$ & 15.57 & 3.66 & 2.18 & 0.49 & 0.05 
& 16 & 2.84 & 2.05 & 1.04 & 0.06 \\
11$^+$ & 0 & $(fp)^{20}g_{9/2}^2$ & 14.86  & 2.53 & 1.44 & 0.70 & 1.92 
& 16 & 2.05 & 0.88 & 1.01 & 2.06 \\
17$^+$ & 0 & $(fp)^{20}g_{9/2}^2$ & 13.81 & 3.25 & 2.40 & 0.16 & 2.14 
& 16 & 1.88 & 2.09 & 0.00 & 2.03 \\
2$^+$ & 1 & $(fp)^{22}$ & 15.48 & 3.85 & 1.70 & 0.73 & 0.08 
& 16 & 3.09 & 1.55 & 1.11 & 0.24 \\
8$^+$ & 1 & $(fp)^{22}$ & 15.58 & 3.87 & 1.89 & 0.47 & 0.06 
& 16 & 2.59 & 2.54 & 0.72 & 0.16 
\\ \hline 
\end{tabular}
}
\end{table}%

The total occupancy in the $fp$-shell orbits $f_{7/2}$, $p_{3/2}$,
$f_{5/2}$, $p_{1/2}$, and in $g_{9/2}$ covers 21.45-21.95 particles in
CNS, and the remaining part, 0.55-0.05 particles, is distributed over
other shells due to $\Delta N$=2 oscillator shell mixing.  It is seen
that the two bands described as $T$=0 (exemplified by $I^{\pi}$=3$^+$
and 9$^+$) and $T$=1 ($I^{\pi}$=2$^+$ and 8$^+$) with the same
configuration, $(fp)^{22}$, have rather similar occupation numbers.
This is also the case in the SM calculation.  In general, the
occupation numbers are quite similar in CNS and¨ SM calculations. For
example, the occupation of $f_{5/2}$ increases with increasing spin in
both models. The major difference between the two calculations is the
relative distribution of particles in $p_{3/2}$ and $p_{1/2}$. All
calculated states in SM have almost 1 particle less occupying
$p_{3/2}$ and 0.5 occupying $p_{1/2}$ more, as compared to the
corresponding states in CNS. This may be related to the difference in
the single-particle spin-orbit energy splitting between $p_{3/2}$ and
$p_{1/2}$ in SM (1.1 MeV) and in CNS (2.8 MeV).

Largest deviations between SM and CNS occupancies appear above
the band-crossing, in particular at the 17$^+$ state
which terminates the rotational band. In the corresponding 
configuration the three ``valence'' protons and neutrons
occupy the $\Omega$=9/2 orbital from $g_{9/2}$, $\Omega$=5/2 orbital from 
$f_{5/2}$ and $\Omega$=3/2 orbital from $p_{3/2}$, giving
$I=2\cdot(9/2+5/2+3/2)=17$. Due to a rather large oblate
deformation of the 17$^+$ state ($\varepsilon \approx 0.27$, 
$\gamma = 60^{\circ}$) predicted in CNS,
the orbitals emerging from $p_{3/2},$ $f_{5/2}$ and $p_{1/2}$ shells
contain large components
of the (spherical) $f_{7/2}$ shell. Due to this mixing
the $17^+$ state contains approximately two holes in 
the spherical $f_{7/2}$ shell, see Table 1.
Note that this band is described as [01,01] in the notation of 
ref.\ \cite{Zn62}, i.e., with no holes in orbitals emerging from the
(deformed) $f_{7/2}$ shell in the CNS calculation.

Thus from the results presented in Table 1, one may see that,
according to the CNS calculation, it might be important to include
$f_{7/2}$ in the SM calculation for an accurate description of states
in the $(fp)^{20}g_{9/2}^2$ band, particularly close to band
termination.

% --------------------------------------

\subsection{Backbending}
Both unpaired CNS calculations and full SM calculations reproduce the
observed backbending very well, see fig.1.  In fact, also ``unpaired''
SM calculation describes the backbending quite well (fig.6a). From the
CNS calculation we find that for $I \geq 9$ it is energetically
favorable to excite the two odd nucleons from the $fp$ shell into
$g_{9/2}$ orbits.  Also the SM calculation suggests that the lowest
$T$=0 states with $I$$\ge$9 are formed by exciting two nucleons to
$g_{9/2}$, see Table 1. The backbending is thus suggested to be caused
by an {\em unpaired band-crossing\/} involving a two-particle-two-hole
excitation. The origin of the backbending in $^{62}$Ga is thus very
different from backbending in heavier nuclei, where it is understood
as rotation-induced alignment of angular momentum vectors of a pair of
neutrons (or protons) in a high-$j$ orbit \cite{backbending}. This
type of backbending is an effect from pairing, and does not appear in
unpaired solutions. Unpaired band-crossings have been observed in the
$A\sim160$ mass region at high spin (where pairing plays a minor role)
\cite{Riley,IR85}.

In unpaired CNS calculations the backbending in $^{62}$Ga originates
from a sharp single-particle level crossing (for neutrons as well as
for protons) between the $fp$-shell orbit and $g_{9/2}$ orbit. At low
spin, when the deformation is close to prolate, the energetically most
favored $g_{9/2}$ orbit can be approximately assigned as the [440 1/2]
Nilsson orbit with signature $\alpha$=1/2.  The $fp$-shell orbit has
mixed $p_{3/2}f_{5/2}$ components ([321 1/2], $\alpha$=1/2 in the
low-spin limit for prolate deformations). The excitation
(band-crossing) thus leaves the filled (deformed) $f_{7/2}$ shell
intact%
\footnote{\ The band is thus different from
a recently observed excited band in $^{58}_{29}$Cu which has also one
neutron and one proton promoted to the $g_{9/2}$ shell, but where the
excitation takes place from deformed orbits emerging from $f_{7/2}$
\cite{Dirk}, giving rise to a larger deformation ($\varepsilon \approx
0.4$). Compared to the discussed band in \up62Ga, the band in \up58Cu
is obtained by removing two protons and two neutrons from the
$f_{7/2}$ shell (in the notation of ref.\ \cite{Zn62}, the
configuration is [21,21]).%
}.  Also this configuration has approximate isospin $T$=0.  In
the single-particle diagram (fig.4) the excitation can be seen in the
$\alpha$=1 configuration as a lifting of one proton as well as one
neutron from the highest-lying filled negative-parity orbital to the
lowest un-occupied positive-parity orbital. At the rotational
frequency for which fig.4 is drawn (corresponding to
$I$$\approx$6$\hbar$), the excitation is energetically unfavored, but
at higher rotational frequency the positive-parity orbital comes down
in energy below the filled negative-parity orbital, and the excitation
is energetically favorable for $I\ge9$. The excitation implies an
increased quadrupole deformation, from $\varepsilon \approx 0.17$ to
$\varepsilon \approx 0.23$, see fig.5.
\begin{figure}[tb]
\centerline{\psfig{figure=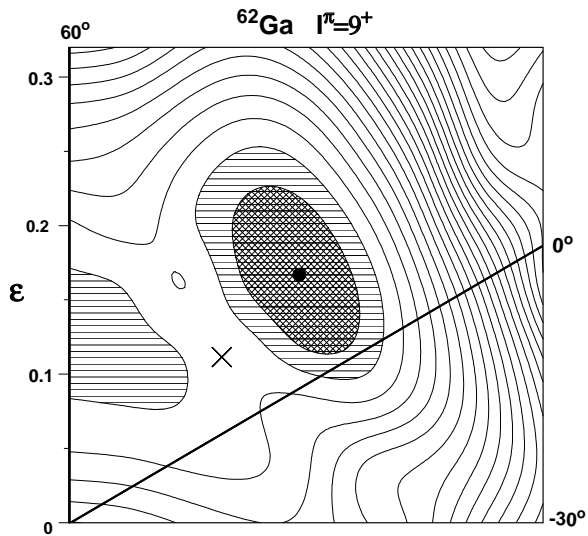,height=\figheight,angle=-90}}
\caption{CNS calculated potential-energy surface in the 
$(\varepsilon,\gamma)$ plane for $^{62}$Ga at $I^{\pi}$=9$^+$. The
line $\gamma$=0$^{\circ}$ corresponds to a prolate shape with
collective rotation, and $\gamma$=60$^{\circ}$ to an oblate shape and
non-collective rotation. The contour line separation is 0.2 MeV. The
filled circle marks the absolute minimum, which corresponds to the
configuration $(fp)^{20}g_{9/2}^2$, while the secondary minimum at
$\varepsilon = 0.13$ and $\gamma = 60^{\circ}$ corresponds to
$(fp)^{22}$. The cross marks the saddle point between the two minima.}
\label{fig7}
\end{figure}
\noindent

At $I$=9 the two configurations come very close in energy, see fig.2.
In CNS the two configurations can be seen as separate energy minima
in the potential-energy surface shown in fig.7. The minimum of the 
$(fp)^{20}g_{9/2}^2$ configuration at ($\varepsilon$,$\gamma$)=
(0.23,$17^{\circ}$) comes 330 keV lower in energy than the $(fp)^{22}$ state
at ($\varepsilon$,$\gamma$)=(0.13,$60^{\circ}$),
and the two minima are separated by a barrier having height of about
500 keV. The 
$(fp)^{20}g_{9/2}^2$ configuration is formed from the $(fp)^{22}$
configuration by exciting one proton and one neutron from the $fp$-shell
to $g_{9/2}$, i.e.\ by a 2p-2h excitation, cf fig.4. Since the single-particle
levels involved in the excitation have different parity they exhibit a sharp
level crossing when plotted versus rotational frequency. Thus, in the mean
field approximation the interaction between the two configurations is
identically zero. The interaction may, however, 
be described beyond the mean field
approximation as a dynamical tunneling phenomenon.
Tunneling between the two minima gives rise to mixing 
of the two wave functions, and two new eigenstates appear.
The SM in principle contains all kinds of dynamical correlations, and
the two 9$^+$ states shown in fig.2 are already mixed.

Backbending is calculated also in the 
$T$=1 and $T$=0 bands with even spins ($\alpha$=0 band in CNS), see fig.6.
Also in these cases the backbending corresponds to a band-crossing
between bands with configurations  $(fp)^{22}$ and $(fp)^{20}g_{9/2}^2$.
In $^{62}$Zn these states have been observed and CNS calculation 
presented in \cite{Zn62}.

% --------------------------------------

\section{Electromagnetic properties}
In previous sections we have shown that the structure of the $T$=0 and $T$=1 
rotational bands changes with increasing spin. The changes occur both
within a given configuration and, at the backbending, from one configuration 
to another. These changes have been seen in both SM and CNS calculations, 
see e.g.\ Table 1. In the CNS calculation large differences were also seen 
between the $\alpha$=1 ($T$=0) and $\alpha$=0 ($T$=0/1) bands in terms of
different kinds of triaxiality (fig.5). These differences are  
not seen in sub-shell occupancies (Table 1) but may be significant 
for other properties.

Electromagnetic properties, such as stretched $B(E2)$ values and spectroscopic 
quadru\-pole moments, are expected to be sensitive to triaxiality and the axis 
around which the rotation occurs. In addition, they are also measurable,
at least in principle. We shall therefore study
electromagnetic properties of the considered states in this section, both
in CNS and SM. Electromagnetic properties were calculated in the SM
using effective charges $e_p$=1.5 and $e_n$=0.5. 
In CNS it is not straightforward to
calculate electromagnetic properties, and in next subsection we describe the
approximative method we used.

% --------------------------------------

\subsection{EM properties in CNS}
Angular momentum is not a good quantum number in the cranking model,
and there is no direct way to calculate electromagnetic
properties. For axially symmetric shapes with collective rotation, the
rotor model can be assumed to be valid, and electromagnetic properties
approximately calculated from the electric quadrupole deformation of
the mean field.  For triaxial shapes the $K$ quantum numbers are mixed
and the simple rotor model cannot be applied.  However, in the
high-spin limit it is possible to derive expressions for $B(E2)$ and
spectroscopic quadrupole moment, $\Qspec$ at any value of the
triaxiality parameter $\gamma$ \cite{BM2}. The combined expressions of
the axially symmetric rotor formulae and the high-spin formulae from
\cite{BM2} were given in \cite{Andrius99}:
\begin{equation}
B(E2; I+2,K \rightarrow I, K)=\frac{5}{6\pi}\,
\langle I+2 K 2 0 | I K\rangle ^2\, Q_2(\hat{x})^2,
\renewcommand{\theequation}{1a}
%\nonumber
\end{equation}
\begin{equation}
\Qspec(I,K)=-2\,\langle I I 2 0 | I I \rangle\, \langle I K 2 0 | I K \rangle\,
Q_0(\hat{x}),
\renewcommand{\theequation}{1b}
\end{equation}
where $Q_0(\hat{x})$ and $Q_2(\hat{x})$ are electric quadrupole
moments around the rotation axis ($x$ axis).  They are calculated in
CNS from the proton wave-functions at the appropriate equilibrium
deformation.  These expressions are thus valid {\em either\/} at
axial-symmetric shapes for any $I$ and $K$, {\em or\/} at triaxial
shapes for high-spin values, when the Clebsch-Gordan coefficients have
taken their asymptotic values (and are independent of $K$). However,
as was seen in ref.\cite{Andrius99}, where $B(E2)$ and $\Qspec$ values
for different states in $^{48}$Cr were calculated from eqs.(1)
and compared to SM results, the expressions seem to work quite well
also outside the region of validity.  With these restrictions in mind we
shall use eqs.(1) to estimate $B(E2)$ and $\Qspec$ values for
different states calculated in $^{62}$Ga.  For $B(E2)$-values the
change of deformation between mother state and daughter state is
neglected; properties of the mother states are simply used.
\begin{figure}[tb]
\centerline{\psfig{figure=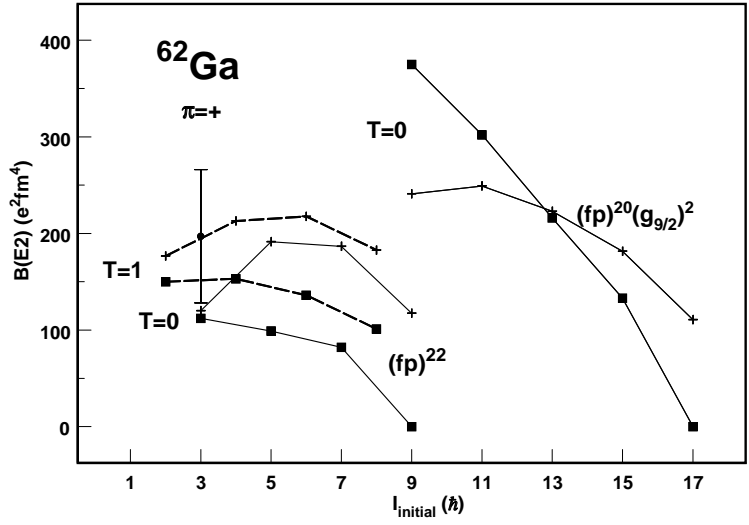,height=\figheight}}
\caption{Calculated $B(E2)$ values in SM and CNS along the $T$=0 band 
(lowest $\alpha$=1 band in CNS), connected by solid lines,
and along the $T$=1 band (lowest $\alpha$=0 band in CNS), connected by 
dashed lines. SM results are shown by plus signs and CNS results with 
filled squares. Approximate configuration below the band-crossing is
$(fp)^{22}$, and $(fp)^{20}g_{9/2}^2$ above it.
Experimental data for the $3^+\rightarrow1^+$ $T$=0 transition  
\cite{Vincent98} is shown by a solid circle with an error bar.}
\label{fig8}
\end{figure}
\begin{figure}[tb]
\centerline{\psfig{figure=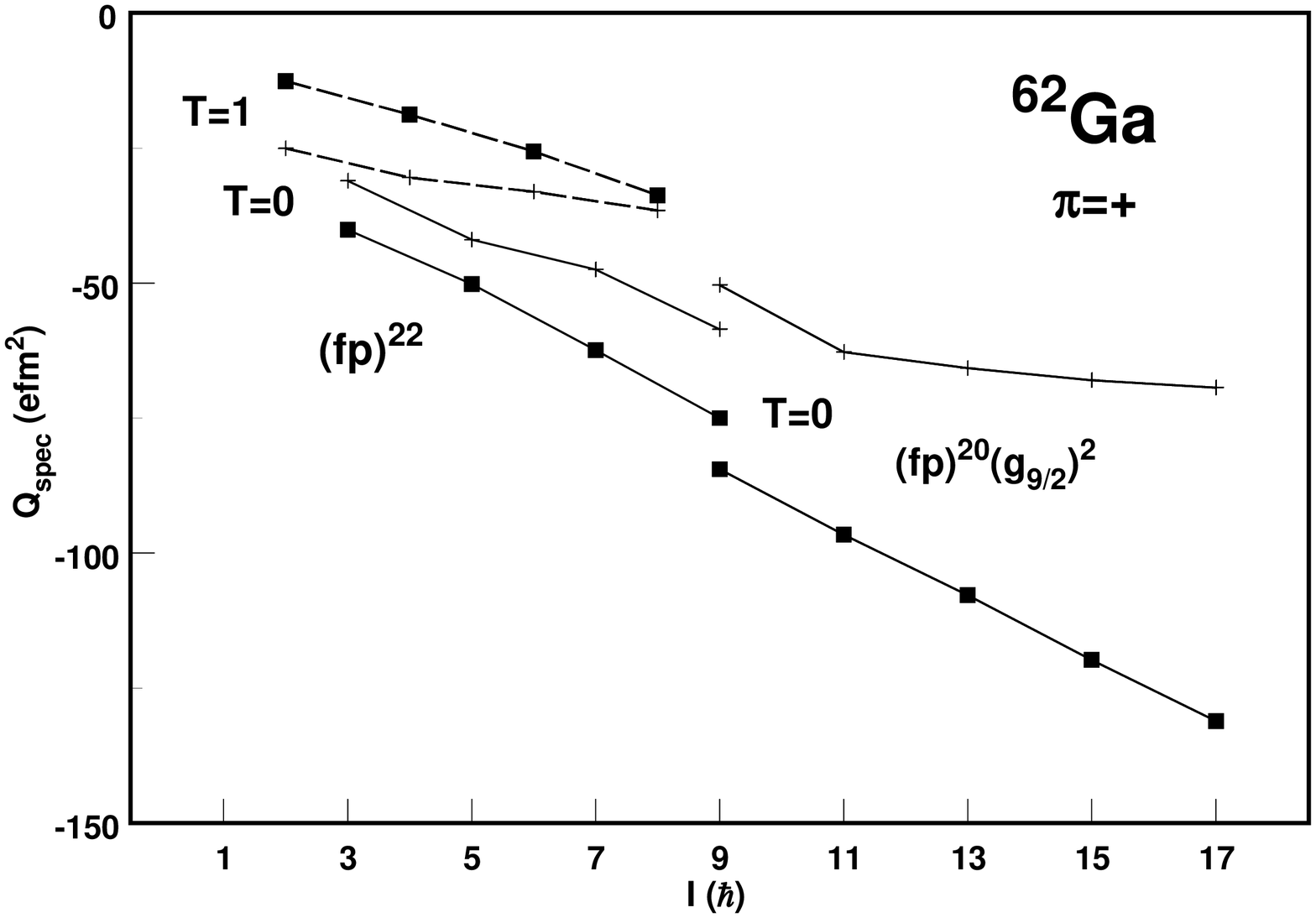,height=\figheight}}
\caption{Calculated $\Qspec$ values in SM and CNS.
See caption to fig.7 for notations.}
\label{fig9}
\end{figure}

% --------------------------------------

\subsection{Comparison between CNS and SM}
In figs.8 and 9 we compare SM with CNS results for $B(E2)$ values and
$\Qspec$ values, respectively.  The $K$ value appearing in eqs.(1) has
been set to $K$=1 for the bands with odd spins, and $K$=0 for the band
with even spins.  The use of eqs.(1) for the low-spin parts of
the bands is outside their expected region of validity, but we still
believe the trends suggested by the calculations contain important
information.  In the CNS calculation the different kinds of
triaxiality of the $(fp)^{22}$ bands with $T$=0 (positive $\gamma$)
and $T$=1 (negative $\gamma$) implies about 50 percent larger $B(E2)$
values for the $T$=1 band compared to the $T$=0 band. In addition,
$|\Qspec|$ is calculated as much larger for states in the $T$=0 band
than in the $T$=1 band.  The increasing $\gamma$ deformation, finally
leading to band termination of the $(fp)^{22}$ band at $I$=9 (fig.5),
implies a gradual loss of collectivity and consequently decreasing
$B(E2)$ values, while $|\Qspec|$ reaches its maximum value. The
different behavior of $B(E2)$ and $\Qspec$ in the two bands 
can be understood from the $\gamma$-dependence of the two
quadrupole moments which appear in eqs.(1): $Q_2(\hat{x})
\propto \cos(\gamma+30^{\circ})$ and $Q_0(\hat{x}) \propto
\sin(\gamma+30^{\circ})$. 

The $B(E2)$ values in the $(fp)^{22}$ bands are calculated as somewhat
larger in the SM than in the CNS model. This might be explained by a
lack of dynamical correlations in CNS. Quantum fluctuations around the
equilibrium deformation, particularly in the $\varepsilon$-direction,
are expected to increase the $B(E2)$ values, see \cite{Andrius99}. The
general trends of $B(E2)$ and $\Qspec$ are, however, similar in the
two models. The difference in $B(E2)$ values and $\Qspec$ values
between the $T$=0 and $T$=1 bands in the $(fp)^{22}$ configuration, as
suggested from the different types of triaxiality obtained in the CNS
calculation, is also seen in the SM calculation. In both models
$B(E2)$ values are larger for the $T$=1 band than for the $T$=0 band
(fig.8), while the opposite is true for $|\Qspec|$ (fig.9). Although
the effect is smaller in SM than in CNS, the CNS view of different
types of triaxiality is supported by the SM results.  Indeed, if
$\varepsilon$ and $\gamma$ deformations are deduced from SM calculated
$B(E2)$ and $\Qspec$ values through eqs.(1), both $T$=0 and
$T$=1 bands are found to correspond to triaxial shapes ($\gamma \neq
0$). Furthermore, the $T$=1 band corresponds to ($\gamma\simleq 0$),
while states in the $T$=0 band has $\gamma > 0$.

The decrease of $B(E2)$ with increasing
spin, particularly seen in the $T$=0 band, is much less accentuated in the SM.
$E2$ transitions along the band with the $(fp)^{20}g_{9/2}^2$
configuration, that is yrast above the band-crossing, are
stronger than for the $(fp)^{22}$ band in both CNS and SM calculations.
But, as discussed in section 3, also states within this configuration 
lose collectivity as angular momentum is increased, and in the CNS
calculation the rotational band terminates
at $I^{\pi}$=17$^+$ in an oblate shape with the symmetry axis 
coinciding with the rotation axis. The alignment sets in smoothly and the
$\gamma$-deformation changes in a gradual way 
from $8^{\circ}$ for the $7^+$ state
to $60^{\circ}$ for the $17^+$ state, see fig.5.
This implies a gradual decrease of $B(E2)$
with increasing spin, as seen in fig.8.
The decreasing behavior of $B(E2)$ in the band above the band-crossing
also appears in the SM results, however, much less drastic than in 
CNS. 

The restricted model space in the SM calculation, particularly the
lack of holes in $f_{7/2}$, is more severe for the description of
states in the $(fp)^{20}g_{9/2}^2$ configuration than in the
$(fp)^{22}$ configuration, especially at higher spins, see Table
1. Effects from the restricted configuration space are seen in
energies (see fig.2), but become much larger for the electromagnetic
properties. Indeed, $|\Qspec|$ is in general considerably smaller in
SM than in CNS.  States within the $(fp)^{20}g_{9/2}^2$ configuration
are calculated in CNS to have $\varepsilon \approx 0.24$ (and $\gamma$
between $10^{\circ}$ and $60^{\circ}$) while corresponding
$\varepsilon$ values in the SM calculation, as deduced from eqs.(1),
are considerably smaller, $\varepsilon \approx 0.16$. This is
explained by the deformation driving effect of holes in $f_{7/2}$ that
is then missing in SM wave functions.

% --------------------------------------

\section{Summary}
The rotational behavior of low-lying
$T$=0 and $T$=1 bands in the odd-odd $N$=$Z$ nucleus $^{62}$Ga
have been studied and compared in the cranked Nilsson-Strutinsky model and
the spherical shell model. 
Both  models have certain advantages and disadvantages that are analyzed in
this comparative study. Their results were also compared with the observed
$T$=0 band in $^{62}$Ga as well $T$=1 band in $^{62}$Zn.

CNS is an approximative mean field theory without dynamical
correlations.  In addition, pairing is neglected in the present
study. On the other hand, CNS provides a good and intuitive picture of
different kinds of collective phenomena, and works well in more or
less all mass regions of nuclei, at all deformations, and in an extended
region of angular momenta \cite{NuclearShapes}.
The SM works very well in restricted areas where the valence
space is not too large. The wave functions are extremely complex and
the intuitive picture of collective phenomena is missing. By comparing
SM and CNS results we have obtained an understanding of different
collective phenomena in $^{62}$Ga, such as rotation,
backbending/band-crossing and band termination.  Furthermore, we have
studied in SM how isoscalar and isovector pairing 
affect $T$=0 and $T$=1 states differently with increasing spin.

Our main conclusions about $^{62}$Ga properties are:
\begin{itemize}

\item The backbending seen in the $T$=0 band of 
$^{62}$Ga is caused by an unpaired band-crossing between two bands with
different configurations. Below the band-crossing all 22 particles
outside the $^{40}$Ca core occupy the $fp$ shell, and above the
band-crossing one proton and one neutron are excited to
$g_{9/2}$. This picture is supported both by CNS and SM calculations,
and both models give energy states for the $T$=0 rotational band
in good agreement with data, see fig.1.

\item Backbending is predicted at $I$$\approx$10 in both $T$=0 and $T$=1 
even spin bands.

\item 
Pairing energy correlations in the (odd spin) $T$=0 band are less
sensitive to increasing spin than in the (even spin) $T$=1 band, see
fig.3. The effect is quite similar for isoscalar and isovector
pairing, although isoscalar pairing is slightly more
stable. Consequently, the moment of inertia is approximately constant
for the $T$=0 band while it increases with increasing spin for the
$T$=1 band. This results in a crossing between the $T$=0 band and
$T$=1 band at $I\approx 2$ seen in fig.2.

\item The blocking of the isovector pairing for $T$=0 states implies 
a smaller (negative) energy contribution from isovector pairing to the
$I$=1 state ($T$=0), than to the $I$=0 state ($T$=1) (fig.3). This 
favors the $I$=0 ($T$=1) state that becomes the ground state.

\item The excitation energy of the $I$=1 ($T$=0) state is well 
reproduced in the SM calculation.

\item Energies of $T$=0 band with odd spins come out very similar in 
unpaired CNS calculation (lowest signature $\alpha$=1 band), and in SM
calculation, when isoscalar and isovector pairing in the $L$=0 channel
has been removed, see fig.6. In a similar way energies in the $T$=1
and $T$=0 bands with even spins come out similar to CNS band with
$\alpha$=0.

\item In CNS both bands involved in the band-crossing show band termination, 
at $I$=9 and $I$=17, respectively. At the band-crossing at $I$=9 the 
ground band terminates its rotational structure, and only one band 
continues after the band-crossing. The termination of the 
$(fp)^{20}g_{9/2}^2$ band at $I=17$ is smooth and favored. 

\item The size of the $E2$ transition matrix element
decreases with increasing spin in both models, see fig.8,
although the effect is weaker in SM.

\item 
From CNS it is suggested that the $T$=0 and $T$=1 bands at low spins
correspond to triaxial shapes with rotation around the intermediate
and the smallest axes, respectively.  This implies quite different
electromagnetic properties for the two bands, see figs.8 and 9, and is
partly supported by SM calculations.

\item At higher spins the used SM configuration space becomes insufficient,
and the need for involving holes in $f_{7/2}$ was stressed.

\end{itemize}

\vspace{1cm}

\noindent
{\bf {\Large Acknowledgments}} \\

We thank Ingemar Ragnarsson and Dirk Rudolph for useful discussions
and comments on the manuscript.  We would like to thank Etienne
Caurier for access to the shell model computer code \cite{Caurier}.
A.\ J.\ thanks the Swedish Institute (``The Visby programme'') for
financial support, and S.\ \AA.\ thanks the Swedish Research Council
(NFR).

\end{document}